\documentclass[12pt]{iopart}
\usepackage{epsfig}
\usepackage{graphics}
\usepackage{amstext}
\usepackage{rotating}

\begin{document}
\title[]{A First Comparison of SLOPE and Other LIGO Burst Event Trigger Generators}

\author{Amber L. Stuver, Lee Samuel Finn}
\address{Center for Gravitational Wave Physics, The Pennsylvania State University, University Park, PA 16802, USA}

\ead{\mailto{stuver@gravity.psu.edu}}

\begin{abstract}
A number of different methods have been proposed to identify unanticipated burst sources of gravitational waves in data arising from LIGO and other gravitational wave detectors. When confronted with such a wide variety of methods one is moved to ask if they are all necessary: i.e., given detector data that is assumed to have no gravitational wave signals present, do they generally identify the same events with the same efficiency, or do they each `see' different things in the detector? Here we consider three different methods, which have been used within the LIGO Scientific Collaboration as part of its search for unanticipated gravitational wave bursts. We find that each of these three different methods developed for identifying candidate gravitational wave burst sources are, in fact, attuned to significantly different features in detector data, suggesting that they may provide largely independent lists of candidate gravitational wave burst events.

\end{abstract}


\section{Introduction}

The great adventure of gravitational wave detection is the search for the unknown: bursts of gravitational waves from sources never anticipated. Searching for the known is a different kind of endeavor than searching for the unknown. In the former case, one is asking an affirmative question of the data--e.g., is a signal like the one I know arises from an inspiraling compact binary system present in the data? In the latter case, however, the best one can do in identifying candidate unanticipated bursts is ask if the data \emph{now} is significantly different than the long-run character of the noise.

The gravitational wave detection community has developed a number of different methods whose goal is to look for such anomalous behavior in the gravitational wave detector data~\cite{anderson, WaveBurst, TFClusters, SP, BN, QPipeline:Shourov}. When confronted with such a wide variety of methods one is moved to ask if they are all necessary: i.e., do they generally identify the same events with the same efficiency, or do they each `see' different things in the detector? The LIGO Scientific Collaboration has used several in its own internal analyses but few burst gravitational wave data analysis method comparisons have been performed.  Two prominent comparisons have been published~\cite{LIGOVIRGO:ETG, Arnaud:Comparison} but neither of these investigated the methods on actual detector data or focused their observations on the results of each method using the current practical application of each method.  Here we consider three different analysis methods--SLOPE \cite{SP}, BlockNormal \cite{BN}, and Q Pipeline \cite{QPipeline:Shourov, QP}--and examine whether they are more or less sensitive to the same types of burst events in the detector data using the current practical tunings of each ETG.

In Section \ref{sec:ETG} we describe the three different analysis methods--SLOPE, BlockNormal and Q Pipeline--that we evaluate in this work. Our discussion of SLOPE is longer than that of BlockNormal or Q Pipeline because references exist for the implementation of these, but not for the implementation of SLOPE that we use here. In Section \ref{sec:NULLBURST} we look in detail at the response of all three to `pure' noise, comparing the false events identified for each and looking for correlations among them.  Finally, we summarize our conclusions in Section \ref{sec:conclusions}.

\section{Several Burst Gravitational Wave Data Analysis Methods} \label{sec:ETG}
Every analysis pipeline contains a component that performs the first identification of what may later become a candidate gravitational wave event. In the LIGO Scientific Collaboration this pipeline component is referred to as an \emph{event trigger generator,} or ETG. While later stages of the pipeline may prune the list of candidate events, it is the ETG that determines the initial list of possibilities. We focus our attention on the three ETGs referred to as SLOPE, BlockNormal and Q Pipeline and the list of triggers each generates. In this section we briefly describe each of these ETGs.

\subsection{SLOPE} \label{sec:SLOPE}

As its name suggests, the SLOPE ETG looks for segments of data that show a mean trend: i.e., that show a statistically significant slope that persists over some period of time. White noise will show, in the mean, no slope, while a gravitational wave burst incident on the detector will show at the very least a rise and fall in amplitude associated with the beginning and end of the burst. SLOPE was originally proposed by Pradier, et al. \cite{SP} and was used internally by LIGO as part of its initial data analysis efforts \cite{S1}.

SLOPE has previously been applied to LIGO data during the first LIGO science data run in 2002.  Ultimately, unfortunate choices in SLOPE's application, as well as immature data conditioning, led to SLOPE being excluded from the burst gravitational wave upper limit calculation \cite{S1}.  Described here are the two major modifications to SLOPE from its first LIGO application\cite{StuverThesis}.

\subsubsection{Thresholding}
In the absence of a gravitational wave signal, assuming that the detector noise is Gaussian and white, the probability of measuring a slope of magnitude greater than $|m|$ is:
\begin{equation} \label{equ:SLOPEprobability}
\Sigma=1-2\int_{0}^{|m^{'}|}\frac{1}{\sqrt{2\pi}}\exp\left(\frac{(|m|-\mu_{m})^{2}}{2\sigma^{2}_{m}}\right)\mathrm{d}m
\end{equation}
where $\mu_{m}$ and $\sigma^{2}_{m}$ (the mean and variance of the slopes produced by SLOPE) are given by Equations 2.4 and 2.5 in \cite{SP}:
\begin{eqnarray} \label{equ:SLOPEmunu}
\sigma^{2}_{m}=\frac{12f_{data}^{2}}{N(N^{2}-1)}\sigma^2_{data} \\
\mu_{m}=0
\end{eqnarray}
and where $\sigma^{2}_{data}$ is the variance of the \emph{data} being analyzed by SLOPE.  These also approximate the data as Gaussian.

\begin{figure} 
\centering
\includegraphics[width=1.0\textwidth]{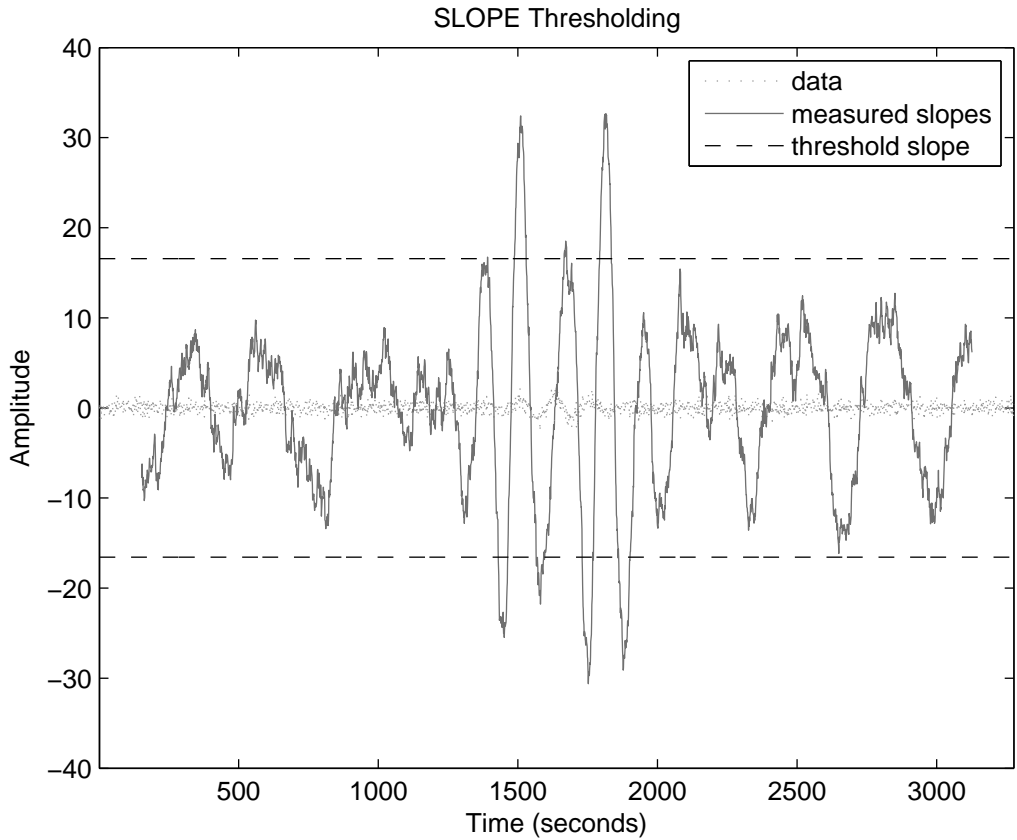}
\caption{Illustration of SLOPE thresholding on a timeseries (dotted line).  Any measured slopes (solid gray line) that are greater than the absolute value of the threshold slope (above the top black dashed line or below the bottom dashed black line) are marked as triggers and later clustered together.}
\label{fig:SLOPEthresh}
\end{figure}

The measured value of $|m|$ is compared against a threshold $m_{thresh}>0$.  $m_{thresh}$ is determined based on the probability $\Sigma$ using the cumulative distribution function $C$:
\begin{equation} \label{equ:cumulativeDistribution}
C(m_{thresh})=\frac{1}{2}\left[1+\mathrm{erf}(\frac{m_{thresh}-\mu_{m}}{\sigma_{m}\sqrt{2}})\right]=\int^{m_{thresh}}_{-\infty}P(m|\sigma_{m}, \mu_{m})dm
\end{equation}
where $C$ is the probability of measuring a slope between $-\infty$ and $m_{thresh}$ and $P$ is the normal probability distribution of slope measurements given $\sigma_{m}$ and $\mu_{m}$.  Since the mean of this distribution is zero (since $\mu_{m}=0$), $P$ is an even function with respect to the measured slopes and comparing the absolute values of the slopes to the threshold slope is justified.  $\Sigma$ is identified as the confidence that a measured slope is not accidental:
\begin{equation} \label{equ:Sigma}
\Sigma=2C-1=\mathrm{erf}\left(\frac{|m_{thresh}|}{\sigma_{m}\sqrt{2}}\right)
\end{equation}
Solving for $|m_{thresh}|$ yields the expression for the threshold slope:
\begin{equation} \label{equ:athresh}
|m_{thresh}|=\sigma_{m}\sqrt{2}\mathrm{erf}^{-1}(\Sigma)
\end{equation}
If for some windowed data $|m|>m_{thresh}$ then a potential gravitational wave candidate is identified.  The value of $m_{thresh}$ is chosen to ensure that the rate of false candidate identifications is an acceptable value.  An illustration of this thresholding method is shown is Figure \ref{fig:SLOPEthresh}.

\subsubsection{Temporal Slope Clustering}

When applying successive windows to the data, a phenomenon we refer to as multi-windowing occurs.  Multi-windowing occurs when a single \emph{event} spans multiple windows, leading to multiple triggers.  Before the SLOPE ETG reports a candidate event triggers that are separated by less than a window duration are clustered together into a single event, whose duration is reported as appropriately greater than a window length.

\subsection{BlockNormal} \label{sec:BlockNormal}

BlockNormal is the name given to an ETG that looks for \emph{change-points} in the statistics of the timeseries data: i.e., points where the mean statistics--in the case of BlockNormal, the data mean and variance--changes. Gravitational waves are uncorrelated with detector noise and, thus, will lead to a change in both mean and variance; so, change-points in the noise statistics mark epochs in which evidence for a gravitational wave burst may be present in the data.  An overview of the entire BlockNormal data analysis pipeline is found in \cite{BN}. 

Once BlockNormal identifies a set of change-points in a timeseries, the mean power in each block (i.e., the epoch between successive change-points) is calculated and compared to a threshold. If the power exceeds the threshold then that block, clustered together with any adjacent blocks whose mean power also exceeds the threshold, is reported as a candidate gravitational wave event.

\subsection{Q Pipeline} \label{sec:QPipeline}

The Q Pipeline ETG is a multi-resolution time-frequency search for excess power in the detector output.  First, the data are whitened using zero phase linear prediction and projected onto bases that are logarithmically spaced in frequency and Q (quality factor), and linearly in time.  Significant tiles whose amplitude significantly exceeds that expected for white, Gaussian noise are reported as triggers assuming white noise statistics and the most significant set of non-overlapping tiles are reported \cite{QP, QPipeline:Shourov}.

\section{Strongest Accidental Event Characterization} \label{sec:NULLBURST}

Our principal objective is to determine the degree to which each ETG sees the data differently. One straightforward way of addressing this question is to run each over the same data set, which may be real detector data or simulated noise, and ask whether the events that are reported by each are the same or different.

\subsection{Methodology and Results}

We ran each ETG over the same set of data, drawn from the LIGO S3 science run, with the thresholds of each set so that they all identified `candidate' events (really, false alarms) at the same rate and with the same mean duration. Each ETG reported back a list of events, each associated with an ETG-dependent amplitude, start-time and duration. These three event lists form the basis for our comparison of the three ETGs. 

The event amplitude reported by each ETG depends on the character of the ETG; however, for identical events the amplitude increases monotonically with the event energy. The events in each list were assigned a rank in their respective list, determined by the reported amplitude of each. If the ETGs are similar we expect that the top-ranked events in each list should be correlated with each other: i.e., the highest ranked events across each list should have similar start times, the second highest ranked events also similar start times, and so on. 

\begin{sidewaysfigure} 
\centering
\includegraphics[width=1.0\textwidth]{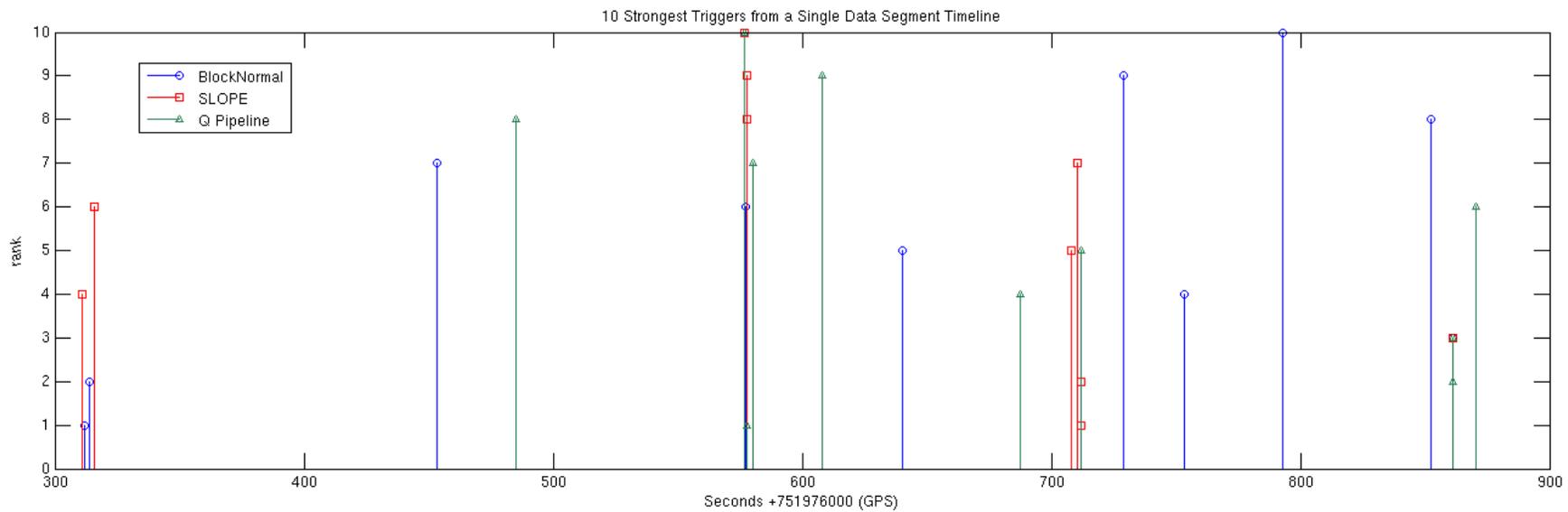}
\caption{Timeline of the top 10 strongest triggers from a single data segment for BlockNormal, SLOPE and Q Pipeline.}
\label{fig:timeline}
\end{sidewaysfigure}

Figure \ref{fig:timeline} shows the start times and amplitudes of the ten strongest events from each ETG from the list generated in one 600 second interval of data. In this segment of data it is clear that some events--for example, the event at GPS time 75198177--are identified by all ETGs; however, the ranks associated with the events range considerably and, while they are closely spaced in time, a more detailed examination of the data shows that there is no overlap of events between the ETGs. This suggests that while there is likely some noise artifact in this area that caused this cluster of triggers, the ETGs did not process it equivalently. Additionally, it is clear that some ETGs identify strong events at places where others do not.

Together, these two observations strongly suggest that each ETG `sees' the data differently.

\subsection{Rank Comparison of Triggers Identified by Multiple ETG's}

Since there were no obvious similarities in the temporal location of the strongest triggers between ETGs and no specific obvious similarities or differences in the timeseries properties upon inspection of the strongest ETG triggers, we looked to see if there was a correlation between the triggers that were detected by ETGs.  Focusing events identified by each ETG in 600 second duration data segments we identified those candidates that were reported by all three ETGs and that shared at least one data sample in common. Each of these triply-coincident triggers was then given three ranks, one associated with its relative amplitude in each ETG. 

There are 253 triply-coincident triggers which are 2.057\% of the triggers admitted by each ETG (top 100 triggers from 123 data sets).  However, there is a multi-triggering issue with Q Pipeline; that is, since Q Pipeline produces single points where the trigger is localized, there is a higher probability that there can be more than one Q Pipeline trigger contained in either a BlockNormal or SLOPE trigger.  There are 233 unique triply-coincident triggers between BlockNormal and SLOPE and 253 unique triply-coincident triggers between BlockNormal and Q Pipeline, and SLOPE and Q Pipeline.  Regardless, the small percentage of triply-coincident triggers compared to the overall number of triggers from each ETG reinforces that there is not an equivalence between the ETG in the triggers that they all identify.

\begin{figure} 
\centering
\includegraphics[width=5in]{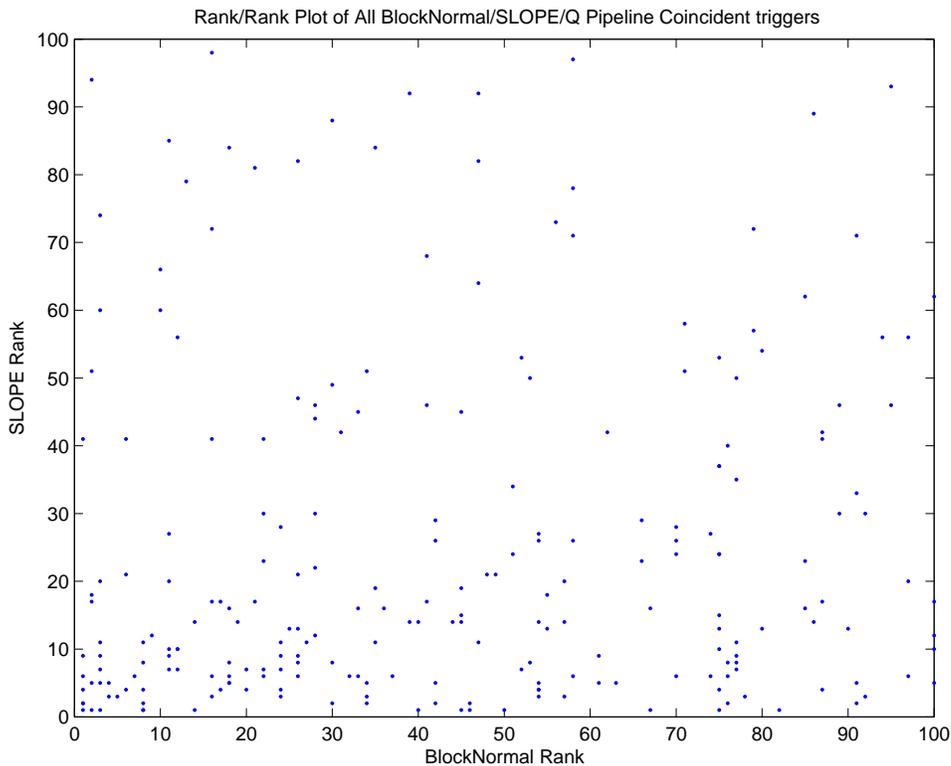}
\caption{Scatter plot showing the rank of a triply coincident trigger in the BlockNormal ETG to the rank as detected in the SLOPE ETG.}
\label{fig:BN&SP3scatter}
\end{figure}

\begin{figure} 
\centering
\includegraphics[width=5in]{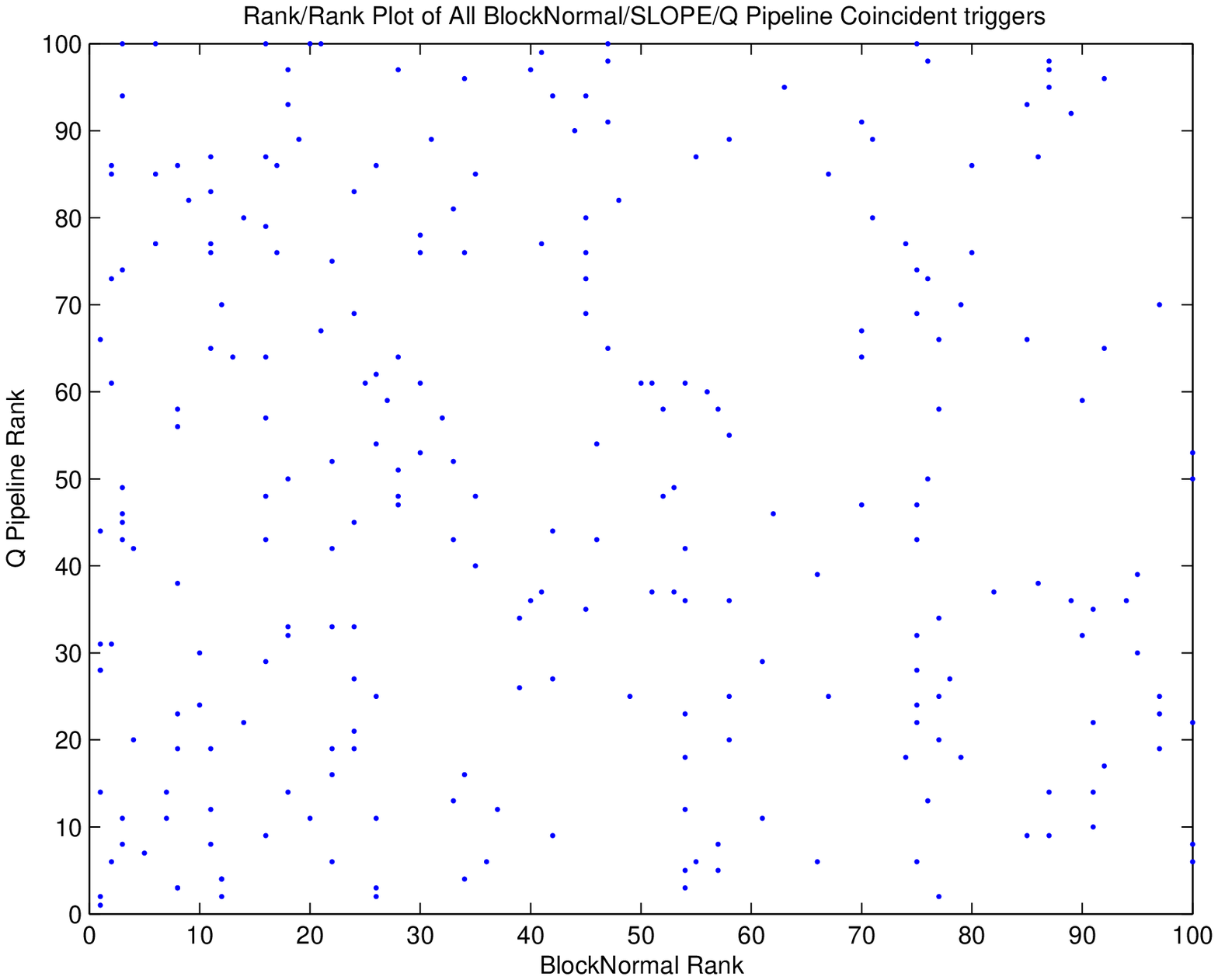}
\caption{Scatter plot showing the rank of a triply coincident trigger in the BlockNormal ETG to the rank as detected in the Q Pipeline ETG.}
\label{fig:BN&Q3scatter}
\end{figure}

\begin{figure} 
\centering
\includegraphics[width=5in]{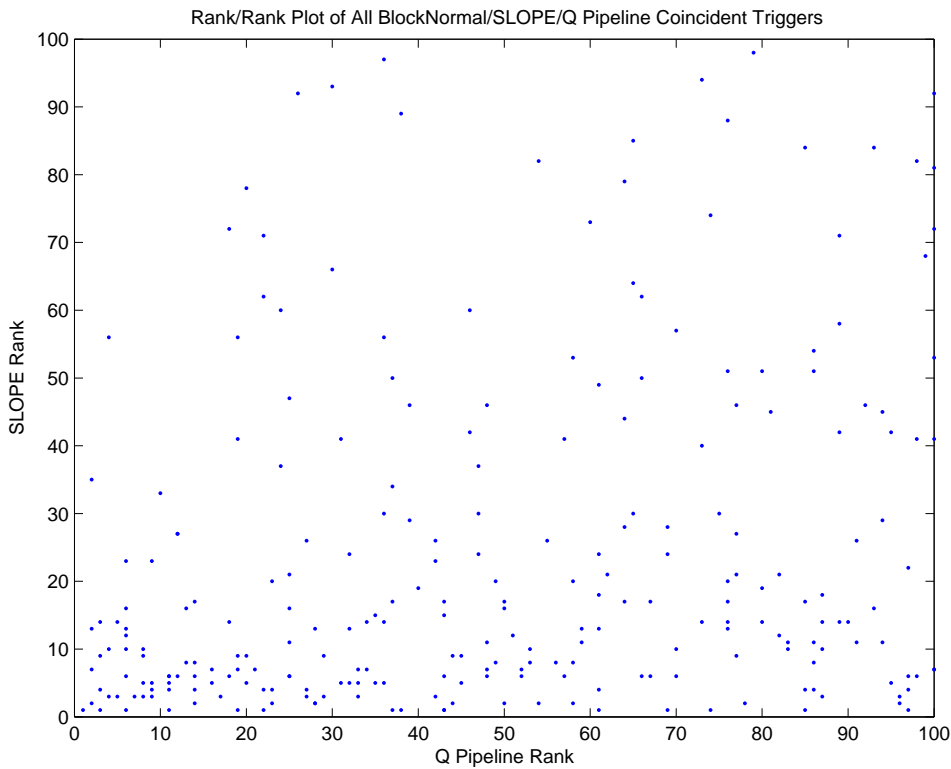}
\caption{Scatter plot showing the rank of a triply coincident trigger in the SLOPE ETG to the rank as detected in the Q Pipeline ETG.}
\label{fig:SP&Q3scatter}
\end{figure}

We then plotted the ranks against each other in a scatter plot, as in Figure \ref{fig:BN&SP3scatter}. If the two ETGs identified events identically, or near identically, we would expect a strong correlation among the ranks, which would show up as a clustering in the scatter plot along a diagonal line radiating from the origin. By eye, Figure \ref{fig:BN&SP3scatter} shows evidence for a very weak correlation in the events identified by BlockNormal and SLOPE, but this is not a correlation in rank equivalence between ETGs. Similarly, Figure \ref{fig:SP&Q3scatter} shows evidence for a somewhat stronger correlation in the events identified by jointly by SLOPE and Q Pipeline, but not in rank equivalence.  Figure \ref{fig:BN&Q3scatter} shows (again, to the eye) no evidence for a correlation among events identified jointly by BlockNormal and Q Pipeline.

\section{Conclusions} \label{sec:conclusions}

This investigation sought to determine if several LIGO burst ETGs revealed similar or different information about the data even though each method processed the data in a different way.  Initial inspection of the temporal locations of the strongest triggers in individual data sets showed that the events identified by each ETG do not form a strongly overlapping set and that, even when we focus just on events that are identified by all three ETGs, each ETG assigns a different relative significance to these events. We conclude that these different ETGs view the data from different perspectives and that candidate gravitational wave events identified in one may be missed by others. Correspondingly, we recommend that multiple methods of searching for gravitational wave bursts continue to be pursued and used in gravitational wave data analysis efforts.

\ack
Special thanks to Shurov Chatterji for helping to incorporate Q Pipeline into this study and to John McNabb for his continuing input on comparison methodology.  We are also grateful to Shantanu Desai, Keith Thorne, and Peter Saulson for comments and useful discussions.

This work was supported by the Center for Gravitational Wave Physics, and the National Science Foundation under award PHY 00-99559. The Center for Gravitational Wave Physics is supported by the National Science Foundation under cooperative agreement PHY 01- 14375.

\end{document}